\documentclass{aa}
\usepackage{graphics}
\begin{document}

\thesaurus{06     
(08.16.4; 08.03.4; 08.13.2; 13.09.6 )} 
\title{2.4 -- 197 $\mu$m spectroscopy of OH/IR stars: \\
The IR characteristics of 
circumstellar dust in O-rich environments
\thanks{Based on observations with {\em ISO}, an ESA project with instruments 
funded by ESA Member States (especially the PI countries: France, Germany, 
the Netherlands and the United Kingdom) with the participation of ISAS and
NASA.}} 
\author{R.J. Sylvester \inst{1} \and F. Kemper \inst{2}
  \and M.J. Barlow \inst{1} \and T. de Jong \inst{2,3}
    \and L.B.F.M. Waters \inst{2,4} 
      \and A.G.G.M. Tielens \inst{5,6} \and A. Omont \inst{7}}

\offprints{F. Kemper (ciska@astro.uva.nl)}
\mail{ciska@astro.uva.nl}

\institute{Department of Physics and Astronomy, University College London,
Gower Street, London WC1E 6BT
\and
Astronomical Institute, University of Amsterdam, Kruislaan 403,
NL-1098 SJ Amsterdam, The Netherlands
\and
SRON Laboratory for Space Research, Sorbonnelaan 2, NL-3584 CA Utrecht, The 
Netherlands
\and
Instituut voor Sterrenkunde, Katholieke Universiteit Leuven, 
Celestijnenlaan 200B, B-3001 Heverlee, Belgium
\and
Kapteijn Institute, University of Groningen, P.O.~Box 800, NL-9700 AV 
Groningen, The Netherlands
\and
SRON Laboratory for Space Research, P.O.~Box 800, NL-9700 AV Groningen,
The Netherlands
\and
Institut d'Astrophysique de Paris. C.N.R.S., 98b bd. Arago, F-75014 Paris,
France}

\date{Received 30 July 1999 / Accepted 1 October 1999}

\titlerunning{2.4 -- 197 $\mu$m spectroscopy of OH/IR stars}

\maketitle

\begin{abstract} 
Infrared spectra of a number of evolved O-rich stars have been obtained with
the Short- and Long- Wavelength spectrometers on board the Infrared Space
Observatory. The very broad wavelength coverage (2.4--197$\mu$m) obtained  by
combining observations made with the two spectrometers includes practically all
of the flux emitted by the sources, and allows us to determine the emission and
absorption features of the dense circumstellar dust shells. Agreement between
the fluxes obtained by the two instruments is generally very good; the largest
discrepancies are probably due to source variability. 
Our sample of oxygen-rich AGB stars exhibits a wealth of 
spectral features due to
crystalline silicates and crystalline water ice in emission and
absorption. In this study a qualitative overview of all features due to
crystalline silicates and water ice in these high mass loss rate objects
is presented. It seems that there is a certain onset value for the mass
loss rate above which these features appear in the spectrum.
Moreover,   
crystalline silicate emission features have been detected for the first
time at wavelengths 
where the amorphous
silicates are still in absorption, implying different spatial distributions for
the two materials. A spherically symmetric and an axi-symmetric geometry
are proposed.

      \keywords{ Stars: AGB and post-AGB -- circumstellar matter -- Stars:
mass-loss -- Infrared: stars }
\end{abstract}

\section{Introduction}
The post-main sequence evolution of stars of low or intermediate mass takes
these stars 
on to the Asymptotic Giant Branch (AGB), where they lose mass at rates of
10$^{-7}$--10$^{-4}$M$_\odot$yr$^{-1}$.
In the circumstellar outflows molecules are formed and dust 
grains condense. The relative abundances of carbon and oxygen in the
star determine the chemical composition of the gas and dust in the outflows.
Oxygen-rich stars produce silicate dust and molecules such as H$_2$O and OH. If
the mass-loss rate is sufficiently high, the dust completely obscures the star
at visible wavelengths, and the object is known as an OH/IR star because of its
strong emission in the infrared (IR), produced by  the dust grains, and in
radio OH lines, due to maser action by OH molecules. See Habing (\cite{habing})
for a detailed review of AGB and OH/IR stars. The optically-thick dust envelopes
of OH/IR stars may be the result of a recent increase in mass-loss rate: the
so-called `superwind' phase (e.g. Justtanont et al.~\cite{justtanont}; 
Delfosse et al.~\cite{delfosse}). Omont et al.~(\cite{omont})
detected the 43- and 60-$\mu$m emission bands of water ice in the KAO
spectra of a number of OH/IR stars, attributing the bands to the
condensation of water molecules onto silicate grain cores in the dense
outflows.

An important result from the Infrared Space Observatory (ISO) 
mission is the detection of emission from
crystalline silicates in the far-IR spectra of many sources, including the
circumstellar environments of young and evolved stars and solar-system comets
(e.g. Waters et al.~\cite{waters96}, \cite{waters99}). Crystalline silicate 
bands have been detected
in OH/IR star spectra (Cami et al.~\cite{cami}) but are not seen in O-rich AGB
stars with low mass-loss rates, suggesting that the abundance of the
crystalline materials is related to the density of the circumstellar matter at
the dust condensation radius (Waters et al.~\cite{waters96}). ISO  has also
detected thermal emission and absorption
by water, in both the gaseous and solid (ice)
phases (e.g. Barlow \cite{barlow}), from O-rich circumstellar environments.

In this paper we present ISO spectra
of seven well-known  OH/IR stars covering a range of mass-loss rates. The
spectrum of the archetypal Mira variable, $o$~Cet, is presented for
comparison. For most of our targets, the spectra cover the complete
2.4--197$\mu$m spectral range of ISO. 
Sect.~2 of the paper describes the observations and data reduction. 
In Sect.~3, 
the spectra are presented and analyzed, with emphasis on the determination
of the continuum and  the features due to ices and silicates. Concluding 
remarks are made in Sect.~4.

\section{Observations}

\begin{table*}
\caption{Observing details for our sources. Included are the revolution number
and on-target time for the SWS and LWS observations, and the scale factor
applied to the LWS observations to bring the spectra into agreement (see text).}
\label{obs}
\begin{tabular}{llccccc}
\hline
Star          & IRAS name  & Date SWS Obs & Date LWS Obs &LWS Factor &SWS Time &LWS Time\\
              &            & JD 2450000+  & JD 2450000+  &           & (sec)   & (sec) \\
\hline
\object{Mira}          & 02168$-$0312  & 489.47    & 633.83       & 1.5 &3454 &1928\\
\object{CRL 2199}      & 18333+0553    & 749.17    & 746.42       &0.97 &1912 &2228\\
\object{WX Psc}        & 01037+1219    & 433.43    & 614.52       & 1.4 &1912 &4704\\
\object{OH104.9+2.4}   & 21177+5936    & 321.29    & 321.28       & 1.0 &1140 &1268\\
\object{OH127.8+0.0}   & 01304+6211    & 825.20    & 651.02       & 1.0 &1912 &1330\\
\object{OH26.5+0.6}    & 18348$-$0526  & 368.16    & 368.13       &1.15 &1912 &1268\\
\object{AFGL 5379}     & 17411$-$3154  & 879.85    & 507.23       & 0.8 &1912 &1614\\
\object{OH32.8$-$0.3}  & 18498$-$0017  & 358.42    & --           & --  &1912 &-- \\
\hline
\end{tabular}
\end{table*}

Seven of our eight sources were observed with both the SWS and LWS instruments,
while of OH32.8$-$0.3 
only the SWS data were useful. Table~\ref{obs} lists the sources
observed, and the JD dates of the observations 
(henceforth we abbreviate the OH/IR star designations to 
OH104.9, OH127.8 etc). Some of our sources (e.g. OH104.9, OH26.5) 
were observed nearly
contemporaneously with the two instruments, while for others, the two spectra
were taken more than 100 days apart. In the past, modelling work (e.g.
Lorenz-Martins \& de Ara\'{u}jo \cite{lorenz}) has been hindered 
to some extent by the
non-simultaneity of NIR photometry and 10-20~$\mu$m spectra (usually IRAS LRS
data). The fact that the SWS spectrum covers both these wavelength regions at a
single epoch will be useful for future modelling of these sources (Kemper
et al, in preparation).

\subsection{SWS}
The 2.38--45.2 $\mu$m part of the spectrum was obtained using the ISO Short
Wavelength Spectrometer (SWS). 
A detailed 
description of the instrument can be 
found in de Graauw et al.~(\cite{degraauw}). 
Our objects were observed in AOT 1 mode, speed 2, except for Mira (speed 3)
and OH104.9 (speed 1).
The spectrum scanned with SWS contains  12
subspectra, that each consist of two scans, one in the direction of  decreasing
wavelength (`up' scan) and one  in the increasing wavelength direction (`down'
scan).  There are small regions of overlap in wavelength between the
subspectra.  Each sub-spectrum is recorded by 12 independent detectors.

The data reduction was performed using the ESA
SWS {\it Interactive Analysis} package (IA$^{\rm 3}$), 
together with the calibration
files available in January 1999, equivalent to pipe-line version 6.0. 
We started from the Standard Processed Data
(SPD) to determine the final spectrum, according to the steps described in 
this session. 

The observations suffer from severe memory effects in the 4.08--12.0~$\mu$m 
and 29.0--45.2~$\mu$m wavelength regions.  It is possible to correct for the
memory effects for the individual detectors, 
using a combined dark-current and memory-effect subtraction
method.  
This method was applied assuming that the flux levels in these wavelength
regions are very high, and treating the memory effect as giving an additive
contribution to the observed signal. We also assumed that the spurious signal
from the memory effect reaches a certain saturation value very quickly  after 
the start of the up scan, and then remains constant throughout the rest of 
the up scan and the entire down scan.  
This memory saturation value is measured immediately after the down scan is
ended, and is subtracted from the up and down scan  measurements. 
The spectral shape of the memory-corrected down scan is now correct; the error
in the spectral shape of the up scan is corrected by fitting a polynomial to
the up scan and adjusting this fit to   
the down scan, without changing the detailed
structure of the spectrum. The order of the applied polynomial fit
differs per subband,
but is chosen to be in agreement with the spectral shape in that band.
For band 1 and 4 we mostly used polynomials of order 1 or 2, for band 2a,
2b and 3 we predominantly used order 2 or 3, and for band 2c higher order
polynomials (up to order 10) were required to adjust the up scan to the 
down scan.

The spectra of some of our objects showed fringes in the 12.0--29.0 $\mu$m 
wavelength region. This was corrected using the defringe procedures
of IA$^{\rm 3}$.

Glitches caused by particle hits on the detector 
were removed by hand. Glitches 
can be easily recognized: they start with a sudden increase in flux level,
followed by a tail which decreases exponentially with time. Any given glitch 
affects only one of the two scans.

The data were further analyzed by shifting all 
spectra of the separate detectors
to a mean value, followed by sigma clipping and rebinning to a resolution of
$\lambda / \Delta \lambda = 600$, which is reasonable for AOT 1 speed 2
observations.

\subsection{LWS}
We obtained  43--197~$\mu$m grating 
spectra using the LWS instrument. Details of
the instrument and its performance can be found in Clegg et al.~(\cite{clegg})
 and Swinyard et al.~(\cite{swinyard}) respectively. The resolution element 
was 0.3~$\mu$m for
the the short-wavelength detectors ($\lambda \leq 93$ $\mu$m) and 0.6~$\mu$m for the
long-wavelength detectors ($\lambda \geq 80$ $\mu$m). Four samples were taken per
resolution element. 
Between 6 and 26 fast grating scans were made of each target, depending on
source brightness and scheduling constraints. Each scan consisted of a single
0.5-sec integration per sample.

The data were reduced using the LWS off-line processing software (version 7.0),
and then averaging the scans after sigma-clipping to remove the
discrepant points caused by cosmic-ray hits. 

For all our  LWS sources, apart from Mira and WX Psc, the Galactic
background FIR emission was strong, and off-source spectra were taken to
enable the background to be subtracted from the  on-source spectrum.  Galactic
background flux levels were only significant for $\lambda \ge 100$~$\mu$m.
The Galactic background
emission is extended compared to the LWS beam, and therefore gives rise to
strong fringing in both the on- and off-source spectra. The
background-subtracted spectra do not show fringing, indicating that the OH/IR
stars are point-like to the LWS, as expected. 

After averaging and background subtraction (if necessary), each observation 
consisted of ten subspectra (one per detector), which were rescaled by small
factors to give the consistent fluxes in regions of overlap, and merged to give
a final spectrum. 

\subsection{Joining the SWS and LWS spectra}
One of the main goals of this article is to study the overall ISO spectra
of the selected objects. Therefore, it is necessary to join the SWS and LWS
spectra in such a way that the flux levels and slopes of the spectra agree for
both LWS and SWS. Differences in the flux levels of the LWS and SWS spectra
are mostly due to flux calibration uncertainties. Although the spectral
shape is very
reliable, the absolute flux calibration uncertainty is 30\% for the SWS at 45 
$\mu$m (Schaeidt et al.~\cite{schaeidt}), and 10-15\% for the LWS at the 
same wavelength (Swinyard et al.~\cite{swin98}).
Therefore, differences
between the flux levels of LWS and SWS 
which are smaller than 33\% are acceptable within the limits of the combined
error bars.

The SWS and LWS spectra were scaled according to their fluxes in the overlap 
region. Generally this resulted in a shift of less than 20\% 
(see Table~\ref{obs}). In the case of Mira and WX Psc, a much larger shift was
required, presumably  due to the large time interval between the SWS and LWS 
observations of these variable stars.

\section{Results}

\begin{figure*}
\resizebox{\hsize}{!}{\includegraphics{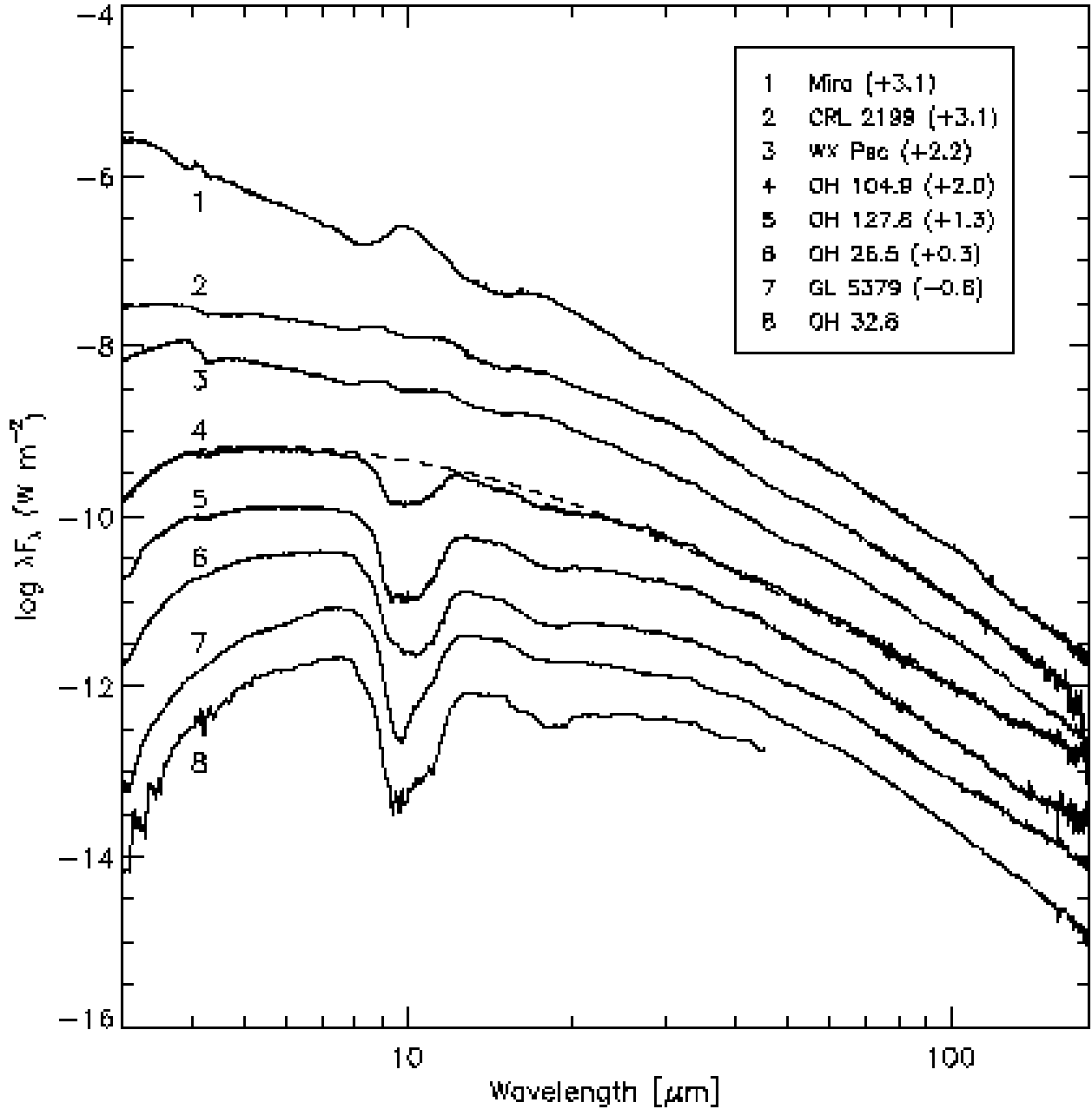}}
\caption{Combined SWS and LWS spectral energy distributions of our programme 
stars. Flux units are
log($\lambda F_\lambda$) in Wm$^{-2}$. The spectra have been shifted along the
ordinate to order them in terms of 10-$\mu$m optical depth, which correlates
roughly with mass-loss rate. The amount by which the spectra were shifted is
given in parentheses. The continuum fit to OH104.9 is also included in the
figure
(dashed line).} 
\label{fig1}
\end{figure*}

\begin{table}
\caption{Dust mass loss rates and SED properties of our sample stars. 
Mass loss rates 
are from Schutte \& Tielens (\cite{schutte}) and 
Justtanont \& Tielens (\cite{justtiel}). $T_{\rm BB}$ is the temperature
of a blackbody which approximates the observed SED. $\tau_{\rm s}$ was measured
directly from the spectra, using continuum fits derived according to the
method described in paragraph 3.1, while $\tau_{mod}$ is the silicate optical
depth derived by 
Justtanont \& Tielens (\cite{justtiel}) from radiative transfer modelling.
}
\label{massloss}
\begin{tabular}{lcrrr}
\hline
Star & $\dot{M}_d$ (M$_{\odot}$ yr$^{-1}$)  &$T_{\rm BB}$(K) &$\tau_{\rm s}$ 
&$\tau_{\rm mod}$\\
\hline
Mira             &                     &1000 &    &\\
CRL~2199         & 1.8$\times 10^{-7}$ &600  &1.2 &3.2\\
WX~Psc           & 7.6$\times 10^{-8}$ &550  &1.3 &\\
OH104.9+2.4      & 2.4$\times 10^{-7}$ &400  &1.2 &\\
OH127.8+0.0      & 2.0$\times 10^{-6}$ &370  &2.2 &16.0\\
OH26.5+0.6       &1.2$\times 10^{-6}$  &350  &2.6 &19.6\\
AFGL~5379        &                     &300  &3.5 &\\
OH32.8$-$0.3     & 2.2$\times 10^{-6}$ &280  &3.9 &19.0\\
\hline
\end{tabular}
\end{table}

The combined spectra are presented in Fig.~\ref{fig1}, 
in $\lambda F_\lambda$ units.

The spectra are ordered by increasing optical depth in the
    observed 10-$\mu$m silicate absorption, 
$\tau_{\rm s}$, and hence are in approximate order of increasing 
mass-loss
rate, assuming roughly similar luminosities. By modelling the infrared excess
emission of O-rich AGB stars, Schutte \& Tielens (\cite{schutte}) and
Justtanont \& Tielens (\cite{justtiel}) have determined the 
dust mass loss rates for several individual O-rich AGB stars.
Their results are summarized in Table~\ref{massloss}, where our sources are
 listed in the same order as in 
Fig.~\ref{fig1}.  
According to Table~\ref{massloss}, our sample is
indeed ordered with increasing mass loss rate, excluding WX Psc.
The values of $\tau_{\rm s}$ measured from the spectra are the
    apparent optical depth compared to our continuum fit to the
    overall SED, or to an assumed silicate emission profile
    for CRL~2199 and WX~Psc. They therefore represent only a part of
the total 10-$\mu$m silicate optical depth towards the sources. This is evident
from Table~\ref{massloss}, where the measured $\tau_{\rm s}$ are quoted, along
with the 10-$\mu$m optical depths derived for some of our sources by 
Justtanont \& Tielens (\cite{justtiel}), using  radiative transfer modelling.
A description of the determination of our continuum fit can be found in
Sect.~3.1.

All of our sources are dominated by continuum emission from cool dust. 
The most striking feature in our spectra is the 10-$\mu$m silicate band, which 
appears in emission for Mira, partially self-absorbed for WX~Psc and CRL~2199,
and in strong absorption for the other sources. 
The overall shape of the observed spectral energy distribution (SED) varies
with $\tau_{\rm s}$, 
the  sources with high $\tau_{\rm s}$ showing `redder' SEDs. The SEDs can be
roughly approximated using blackbodies with temperatures ranging from 300~K for the
most optically-thick sources, to 600~K for 
the partially self-absorbed sources (see Table~\ref{massloss}).

The 20-$\mu$m silicate feature also 
passes from emission to absorption going down
the sequence in Fig.~\ref{fig1}, but none of our sources show it as a
self-absorbed emission feature.
Variations in shape as well as 
optical depth are evident in the 10- and 20-$\mu$m
absorption features. Weaker features beyond 20~$\mu$m can be discerned: these
are features of water ice and crystalline silicates, and will be discussed
below. Evident at short wavelengths are molecular absorption bands (see
Justtanont et al.~\cite{justtanont2} 
for identifications for the supergiant source NML~Cyg,
which is not included here) and
the 3.1-$\mu$m H$_2$O ice absorption feature.

\subsection{Determination of the continuum}

In order to determine the shape and relative strength
of the emission and absorption features we will define a {\it pseudo-
continuum}, which is assumed to represent featureless thermal emission from 
the dust.
The continuum-divided spectrum will provide us with information on 
the optical
depth of the different species that are located outside the continuum-producing
region and the wavelength at which the material becomes optically thin.
However, one has to be very careful with the physical meaning of the
{\it pseudo-continuum}, since the observed continuum results from 
several wavelength-dependent dust emissivities, 
with a large range of dust temperatures, modified by strong optical depth
effects, rather than being simply a superposition of blackbody
energy distributions corresponding to the physical temperature gradient.

For the determination of the continuum we plotted the energy distribution
$\textrm{log}$ $ F_{\lambda}$ 
(W m$^{-2}$ $\mu$m$^{-1}$) against $\lambda$ ($\mu$m). 
Plotting the data this way one can easily recognize the general shape of the
continuum;  
the parts of the spectrum where the continuum is well defined, i.e.~the
long and short wavelength edges, are emphasized. 
This makes it easier to constrain the 
continuum in the wavelength
regions where the strong spectral features are present.
At $\lambda > 50$ 
$\mu$m, the dust is optically thin, and only weak emission features are
present. At $\lambda < 7$ $\mu$m, the radiation from the stellar photosphere
is partially (for the Miras) or completely (OH/IR stars) 
absorbed by molecular lines and the high dust opacity towards the central
star.
When a spline fit 
is performed in  $\textrm{log}$ $F_{\lambda}$ space,
we can thus constrain the continuum by using the known long and short
wavelength continuum contributions as reference points. Applying 
spline fitting with the same reference points 
in other parameter spaces ($\lambda$ $\textrm{log}$ 
$F_{\lambda}$ space, $F_{\nu}$ space, $\textrm{log}$ $F_{\nu}$ space etc.)
shows that all the fits vary within 10\% (in flux) with respect to the adopted 
pseudo-continuum in the 10--20~$\mu$m region.

\subsection{Description of the spectrum}

\begin{figure*}
\resizebox{\hsize}{!}{\includegraphics{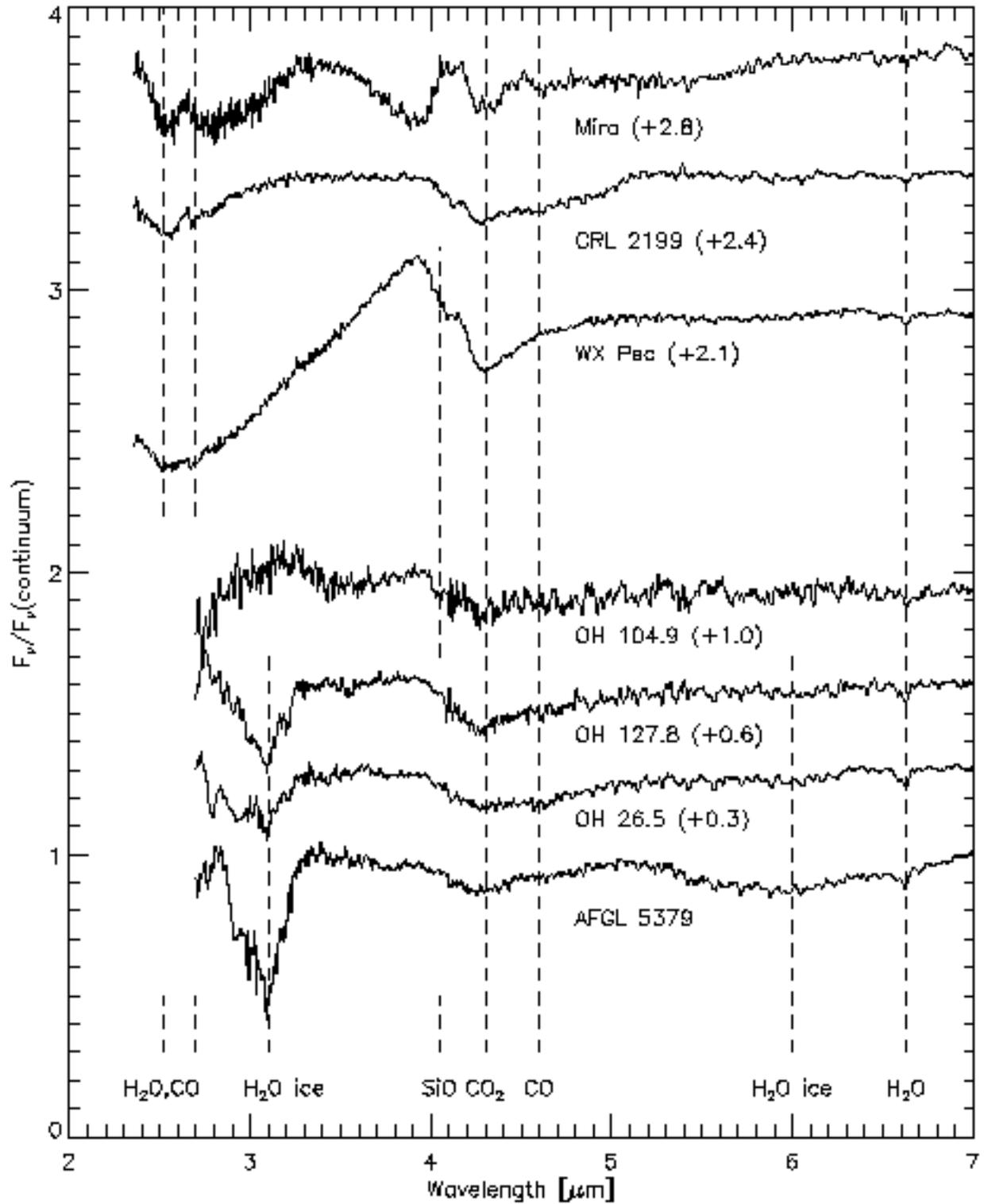}}
\caption{Continuum-divided spectra of our sources in the 2--7$\mu$m region. The
spectrum of OH32.8 is noisy at short wavelengths and has been omitted 
from this figure. The central wavelengths of 
absorption features due to H$_2$O (gaseous and ice phases),
and of gaseous CO, CO$_2$ and SiO are indicated. Typically, the noise level
(standard deviation) of the spectra
is 2\%, but is slightly larger for the short wavelengths of 
the high mass loss objects.}
\label{nir}
\end{figure*}

\begin{figure*}
\resizebox{\hsize}{!}{\includegraphics{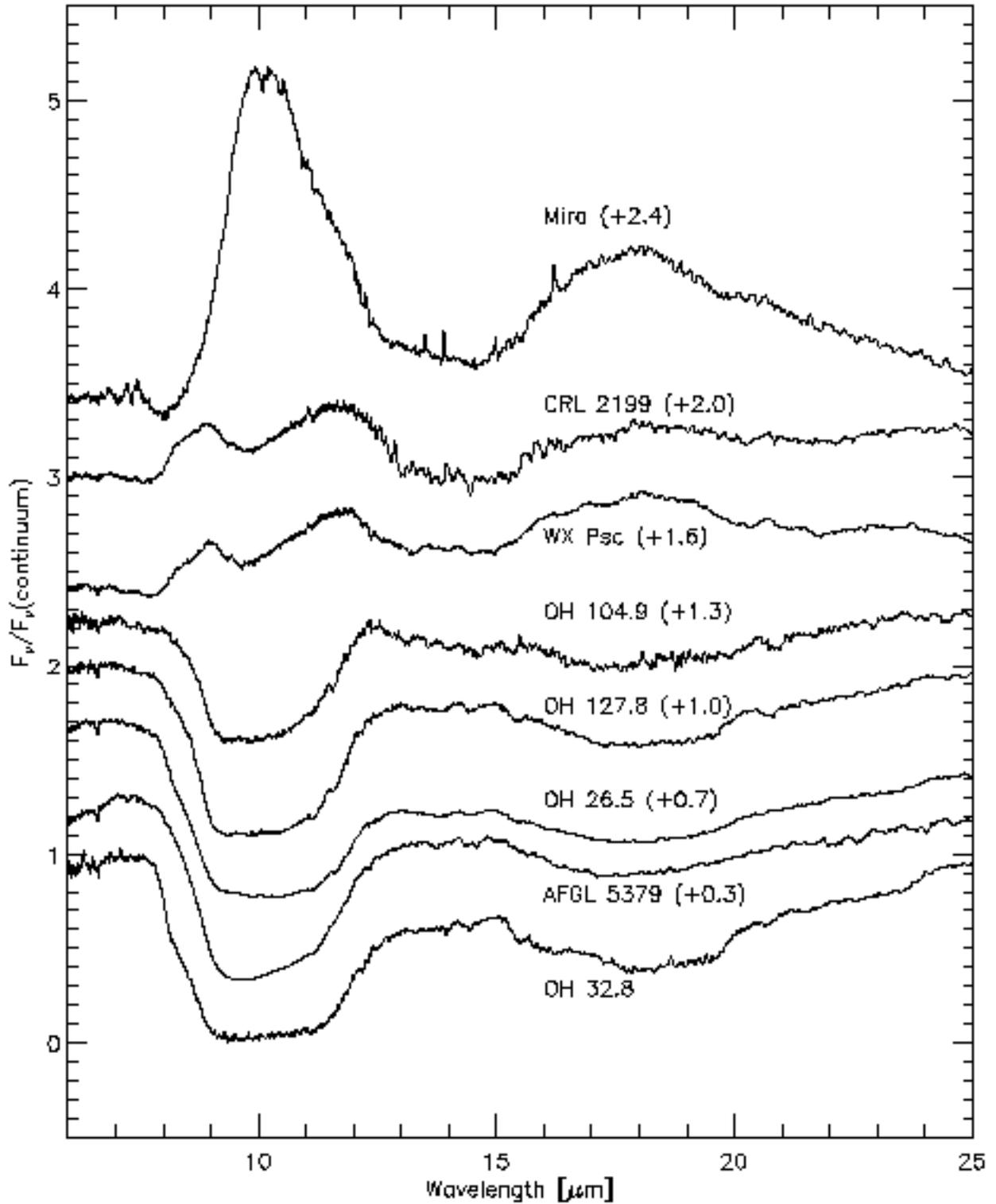}}
\caption{Continuum-divided spectra of our sources in the 6--25$\mu$m region.
The broad amorphous silicate features at 9.7 $\mu$m and 18 $\mu$m 
dominate this region of the spectra. Also visible are the SiO absorption
at 8 $\mu$m and the CO$_2$ emission lines around 15$\mu$m, the latter only
in the spectrum of Mira.
The noise level (standard deviation) 
is typically 1\%, therefore we believe that most of the fine structure visible
superposed on the broad silicate features is real.}
\label{mir}
\end{figure*}

\begin{figure*}
\resizebox{\hsize}{!}{\includegraphics{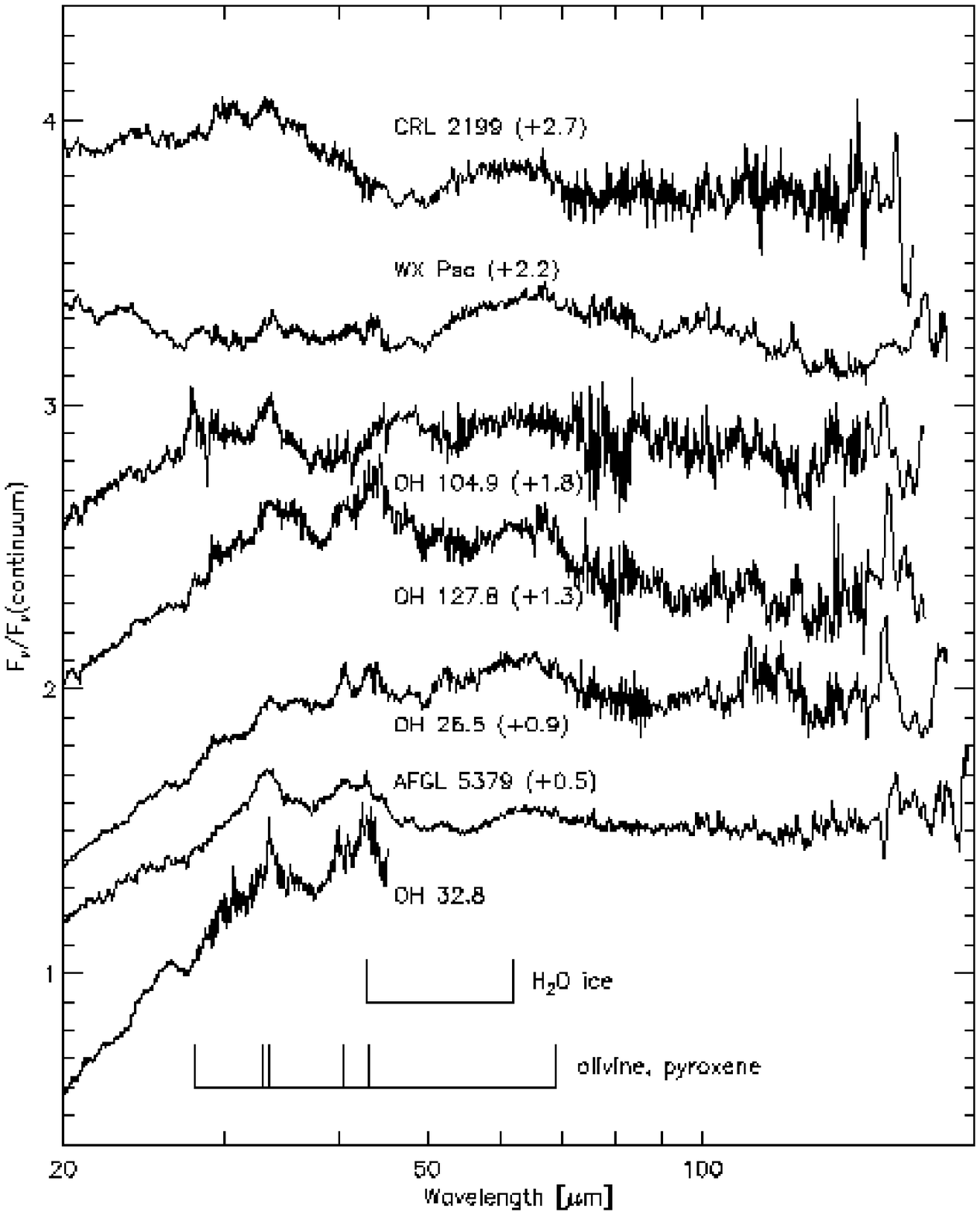}}
\caption{Continuum-divided spectra of our sources in the 20--198$\mu$m region.
The central wavelengths of the broad emission features due to 
crystalline silicates and water ice are indicated.
The noise level (standard deviation) 
of the data is about 2\% up to 45 $\mu$m and about 4\% at
larger wavelengths. The
spectrum of Mira shows poor agreement between the slopes of the SWS and LWS
components, making continuum placement difficult in this wavelength region; it
has therefore been omitted from this plot. There is  no clear evidence for 
crystalline silicates or water ice in our Mira spectrum.}
\label{fir}
\end{figure*}

Figs.~\ref{nir}--\ref{fir} show the observed spectra after division by the
adopted continuum.
A wealth of interesting features can be recognized in the spectra. 
At the shortest wavelengths, from $\lambda = $ 2.38 -- $\sim$7 $\mu$m,
see Fig.~\ref{nir}, the
spectrum is dominated by molecular absorption bands and the
effects of dust absorption. 
In Mira and the intermediate type stars (WX Psc and CRL 2199), strong
absorption features due to a blend of H$_2$O and CO
are present from 2.38--$\sim$3.3 $\mu$m (2.38--$\sim$3.8 $\mu$m
in the case of WX Psc), see Fig.~\ref{nir}.
Absorption due to SiO is found in some objects 
around 4.0 $\mu$m, most prominently in WX Psc and OH104.9. 
In most of our spectra we find clear absorption features of
gaseous CO$_2$ at 4.3 $\mu$m.  CO
line absorption around 4.5 $\mu$m is found in all our spectra.
Throughout the whole 2.38--7 $\mu$m wavelength region, H$_2$O absorption 
lines are present;
very strong lines of gaseous H$_2$O are found at 6.60--6.63 $\mu$m.
Yamamura et al.~(\cite{yamamura}) have performed a detailed study of
the water features in the spectrum of Mira; Mira is also the only source
in our sample that shows OH absorption lines in the 2.5--3.5 $\mu$m region.
In the high mass-loss rate objects, water ice
features are present as well, at 3.1 and
(possibly) 6.0~$\mu$m. The ice features will be discussed in the next section.

The 7--25 $\mu$m region (Fig.~\ref{mir}) 
is dominated by the strong features of
amorphous
silicates. The 9.7-$\mu$m feature (ranging from 8 to 12--13 $\mu$m) and the
18 $\mu$m feature (ranging from  15 to 20--25 $\mu$m) are very strong bands
that occur in emission or (self-)absorption. Crystalline silicate
features also occur in the same wavelength ranges but are usually much weaker, 
so the actual shape of the observed 
silicate bands is due to a blend of crystalline and amorphous silicates.
Absorption by molecular SiO can be seen at 8~$\mu$m, on the wing of the
silicate feature. Around 15 $\mu$m the spectrum of 
Mira exhibits some sharp emission lines
due to CO$_2$.

Longwards of 25~$\mu$m (Fig.~\ref{fir}), most spectral features  
occur in emission. The only absorption feature is the OH pumping line at
34.6 $\mu$m, which is detected in some of the sample stars.
Groups of crystalline silicate emission 
features occur near 28, 33, and 43~$\mu$m, the
latter probably being a blend with the 43 $\mu$m crystalline water ice 
feature, 
while a broad feature, which we ascribe to water ice, peaks around 62~$\mu$m.
Superposed on this is a sharp feature at 69~$\mu$m, due to forsterite.

Several sources 
show emission features near 47.5 and 52~$\mu$m.
The longitudinal optical band of water ice lies close to 52~$\mu$m
(Bertie \& Whalley \cite{bertie}), but in laboratory data it is only 
ever seen as a shoulder on the 
43-$\mu$m band (e.g. Smith et al.~\cite{smithal}), not a clearly-separated
feature. The observed feature is therefore unlikely to be the longitudinal ice
band. 
Malfait et al.~(\cite{malf99}) 
suggested that montmorillonite, which gave a good
fit to the broad 100-$\mu$m emission feature in the spectrum of HD~142527,
is a
possible carrier of the 47 and 50$\mu$m  features in this source. The planetary
nebula NGC~6302 also shows the 47, 53 and 100~$\mu$m bands (Lim et
al., 1999).
If the three features do have a common carrier, the apparent absence of the
100~$\mu$m band in the spectra presented here may be due to a difference
in dust temperature.

As well as broad emission bands, unresolved emission lines can be detected in
the long-wavelength regions of the least noisy of our spectra, (e.g. WX~Psc and
AFGL 5379). Most of these lines are pure rotational lines of water vapour. The
157.7-$\mu$m [C~{\sc ii}] line is also visible in most of our sources;
this is
the residual Galactic background [C~{\sc ii}] emission after subtraction
of the
off-source spectrum. In the case of AFGL~5379, the [C~{\sc ii}] line is
seemingly in
absorption: again, this is due to imperfect cancellation of the background
emission.

\subsection{Water Ice}

Water ice is an important component of the solid-phase material in cool
astronomical sources. Its spectrum shows bands at 3.1, 6.0, 11-12, 43 and 
62~$\mu$m.
The 3.1-$\mu$m stretching band is seen (always in
absorption) in the spectra of many highly-embedded young stars (e.g. Whittet et
al.~\cite{whittet}) and in some OH/IR stars (e.g. Meyer et al.~\cite{meyer}). 
Its formation in the 
circumstellar envelopes of OH/IR stars has been discussed in
particular by Jura \& Morris (\cite{jura}).
Before the ISO
mission, the far-IR ice bands had been observed in emission in a small number
of sources including the OH/IR stars OH26.5, OH127.8 and
OH231.8$+$4.2 (Omont et al.~\cite{omont} and references therein). 
ISO spectra have
shown the 43- and 62-$\mu$m ice bands in emission in various objects (Barlow
\cite{barlow}), such as the
planetary nebulae CPD$-56^{\rm o} 8032$ 
(Cohen et al.~\cite{cohen}) and NGC~6302
(Lim et al.~\cite{lim}), the post Red Supergiant source
AFGL~4106 (Molster et al.~\cite{molster}) and Herbig Ae/Be stars (Waters \& 
Waelkens \cite{wawa};
Malfait et al.~\cite{malfait}, \cite{malf99}),
while the 43-$\mu$m band has
been detected in absorption toward the highly-embedded sources AFGL~7009 and
IRAS 19110+1045 (Dartois et al.~\cite{dartois}, Cox \& Roelfsema \cite{cox}). 
Both of the far-IR bands can be
blended with crystalline silicate emission, but examination of the shapes and
positions of the observed bands can distinguish between silicate and ice
emission.

\begin{figure*}
\resizebox{\hsize}{!}{\includegraphics{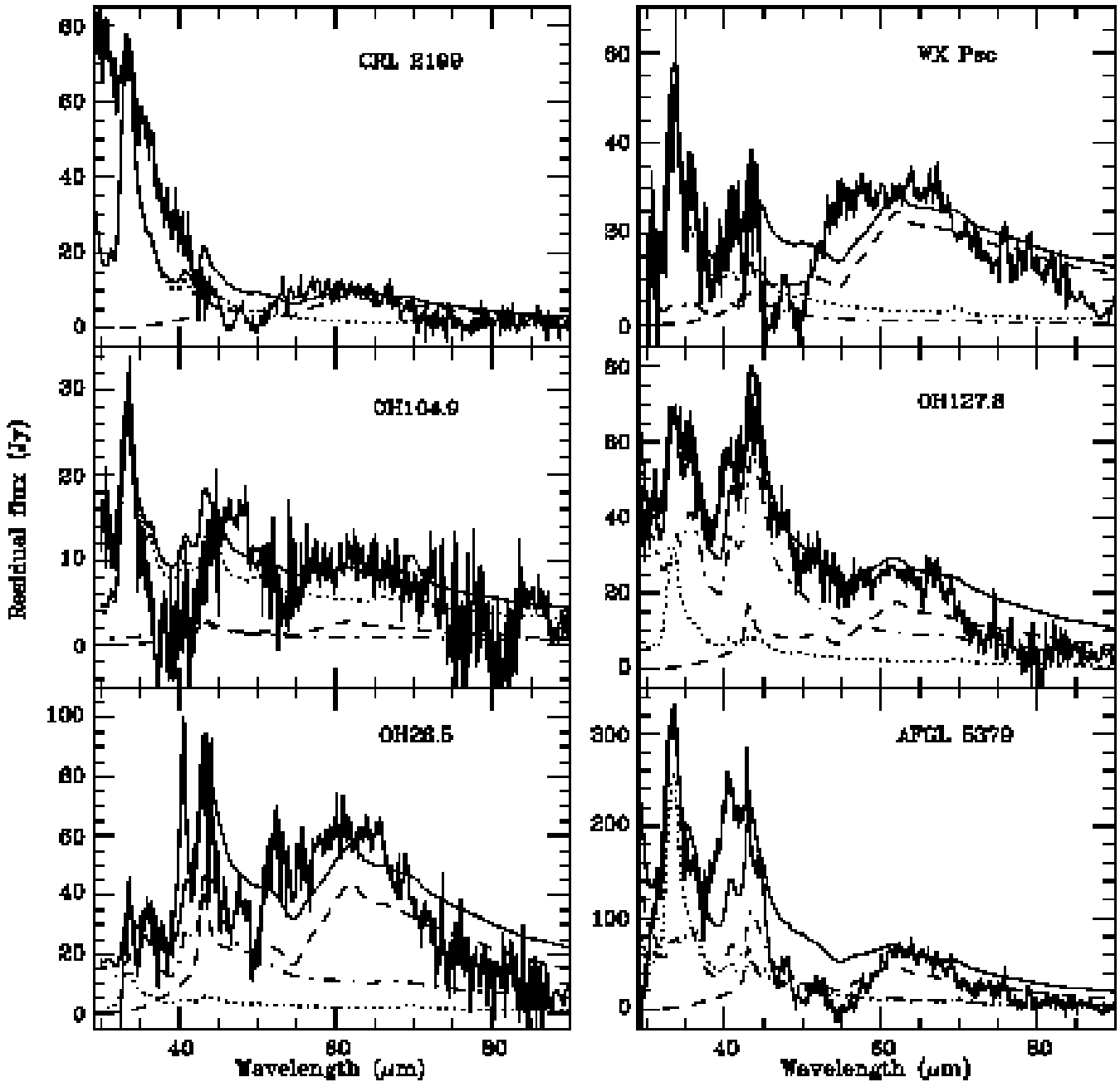}}
\caption{Fits to the spectral features in the 30--90~$\mu$m region. Noisy line:
continuum-subtracted spectrum; dashed line: 
crystalline water ice; dotted line: forsterite; dash-dotted line:
enstatites
(sum of clino- and ortho- forms); solid line: sum of all components.
The optical constants for ice are from Schmitt et al.~(\cite{schmitt}) and
those of the silicates are from J\"{a}ger et al.~(\cite{jaeger}). The observed
43-$\mu$m
features are likely to be blends of ice and pyroxene emission.
} 
\label{icefig}
\end{figure*}

\begin{table}
\caption{Detections of water ice features in the spectra of our programme
stars. $\tau_{3.1}$
and $N_{\rm ice}$ are the optical depth of the 3.1-$\mu$m feature and the
derived column density}
\label{icetab}
\begin{tabular}{l@{\hspace{1mm}}c@{\hspace{2mm}}c@{\hspace{2mm}}c@{\hspace{2mm}}r@{\hspace{1mm}}c}
\hline
Source & Absorption &\multicolumn{2}{c}{Emission} 
  &$\tau_{3.1}$ &$N_{\rm ice}$\\ \hline
&3.1$\mu$m &43$\mu$m &62$\mu$m &&{(10$^{16}$cm$^{-2}$)}\\
Mira     &N &N &N &$\leq$0.02&$\leq$3.4\\
CRL~2199 &N &N &N &$\leq$0.02&$\leq$3.4\\
WX~Psc   &N &Y &N &$\leq$0.02&$\leq$3.4\\
OH104.9  &N &N &? &$\leq$0.05&$\leq$8.4\\
OH127.8  &Y &Y &Y &0.33      &55\\
OH26.5   &Y &Y &Y &0.25      &42\\
AFGL~5379  &Y &Y &Y &0.67    &112\\
OH32.8 &Y$^1$ &Y &-- &0.7$^2$     &120$^2$\\ \hline
\end{tabular}

$^1$ Roche \& Aitken (\cite{roche}) \\
$^2$ Meyer et al (\cite{meyer})
\end{table}
 The new ISO spectra have significantly better resolution and sensitivity
than the earlier KAO data.  Fig.~\ref{icefig} shows attempts to fit the
30--90~$\mu$m region of our (continuum-subtracted) spectra, using a spectral
synthesis routine kindly provided by Dr T.~Lim (personal comm.). This routine
takes absorption (or emission) efficiencies for materials of interest, along
with user-defined temperatures and relative amounts, and produces the
resulting spectrum for optically-thin emission from the individual materials,
as well the total emission from all the materials (shown as the thin solid
line in Fig.~\ref{icefig}). The materials used to fit the OH/IR star spectra
were forsterite, enstatite and crystalline water ice; temperatures of order
50--100~K were used for the fitting.
The detected ice 
features are listed in Table~\ref{icetab}.

Pyroxenes (such as enstatite; see dash-dotted line in Fig.~\ref{icefig})
 also show strong 43-$\mu$m features, therefore detection of a
43-$\mu$m band in an observed spectrum is not sufficient evidence to
demonstrate the presence 
of H$_2$O ice. However, for temperatures $\ga$40~K, the
43-$\mu$m peak is at least as prominent as the 62-$\mu$m peak, so objects which
do not have a 43-$\mu$m feature are unlikely to contain much water ice (unless
it is very cold). Conversely, if an object shows the 62-$\mu$m feature, H$_2$O
ice is likely to be responsible for at least part of that object's 43-$\mu$m
feature.

We claim detections of crystalline water ice emission based on the presence of
the bands at 43 and 62~$\mu$m in OH127.8, OH26.5 and AFGL~5379, confirming the
tentative detections for the first two sources by Omont et al.~(\cite{omont}).
OH32.8 also shows a 43-$\mu$m feature (Fig.~\ref{fir}), 
but without observations
of the 62-$\mu$m feature, we cannot determine if ice emission is present.

CRL~2199 and WX~Psc both seem to show broad 50--70~$\mu$m features, but 
the shape
of these features does not resemble laboratory crystalline ice features, 
unlike the observed features of OH127.8, OH26.5 and GL5379. In particular, the
emissivity of ice has a minimum near 55~$\mu$m before reaching its peak at
62~$\mu$m (see dashed line in Fig.~\ref{icefig}). 
This structure is present in the spectra of the latter three
sources, while the CRL~2199 and WX~Psc features are more flat-topped, with
strong emission at 55~$\mu$m  and no real evidence of a peak at 62~$\mu$m. The
attempt to fit the WX~Psc spectrum with ice emission (Fig.~\ref{icefig})
illustrates this point. The OH~104.9 features are rather weak and ill-defined;
there may be a weak 62-$\mu$m band, but the 43-$\mu$m band is replaced by a
broader feature peaking near 47~$\mu$m.

If the 50--70~$\mu$m features in CRL~2199 
and WX~Psc are not water ice, what are
they? They may simply be instrumental artefacts: Fig.~\ref{fig1} illustrates
how the spectra are dominated by the steep downward slope of the SED, and that
spectral features around 60~$\mu$m are small perturbations on this overall
trend. (Clearly, this statement also holds for the features which we believe to
be real.) Experiments with adopting different LWS dark currents and with
methods for dealing with detector memory effects made little difference to the
spectra. The clino-pyroxene optical constants of Koike et al.~(\cite{koike}) 
show a
broad feature around 60~$\mu$m (see also Cohen et al.~\cite{cohen}), but this
peaks at longer wavelengths than the crystalline water ice band, and so is not
a likely carrier for the CRL~2199 and WX~Psc features. 
Similarly, the 62-$\mu$m features in OH127.8, OH26.5 and GL~5379 do not need an
additional long-wavelength component, so the Koike pyroxene is not necessary to
fit these spectra. The J\"ager et al.~(\cite{jaeger}) laboratory data do not
reproduce the broad 60-$\mu$m band seen in the Koike data, so the feature may
not be real.

In general, we find that the observed features in Fig.~\ref{icefig}
are narrower than our fits, suggesting that we have overestimated the
continuum level in this wavelength region, or that the optical constants we
used do not adequately represent the circumstellar materials.

The sources in which we detect the ice emission features are the four
stars
with the deepest 10-$\mu$m silicate absorption,
and hence presumably the highest
mass-loss rates. These same stars also show the
3.1-$\mu$m absorption band (see Fig.~\ref{nir}).
OH32.8 was too faint at 3~$\mu$m to be detected by the SWS,
but the 3.1-$\mu$m ice absorption feature has been detected in
ground-based spectra (Roche \& Aitken \cite{roche}).
The sources
with self-absorbed silicate emission features, do not appear to show ice
features. OH104.9, which shows a relatively shallow silicate absorption,
may show
a weak 3.1-$\mu$m absorption, but it is hard to discern, because the
spectrum
is noisy at short wavelengths.

The 6.0-$\mu$m band of water ice is significantly weaker than the 3.1$\mu$m
band (see e.g. Moore \cite{moore}). The spectra of our most heavily-obscured
sources, OH32.8, AFGL~5379 and OH26.5, show a weak depression around
6~$\mu$m,
but this wavelength region is very rich in gaseous H$_2$O lines, making it
difficult to ascribe the observed feature to ice absorption.

Ice formation requires cool temperatures and sufficient shielding from
stellar and
interstellar radiation (e.g. Whittet et al.~\cite{whittet}). The high
densities in the (general) outflow that accompany large mass-loss rates 
may provide the required shielding. Alternatively, enhanced densities could be
provided by the formation of a circumstellar disk in the superwind phase, or by
inhomogeneous mass loss, such as is apparent in studies of H$_2$O and OH
maser clumps (e.g. Richards et al.~\cite{richards}). As discussed by
Omont et al.~(\cite{omont}), the presence of the 63-$\mu$m band requires 
that the water ice is at least partially crystalline, implying that the ice
remained relatively warm ($\ga$100~K) for long enough to allow crystalline
reorganization to take place.

 Optical depths and column densities for the detected
3.1-$\mu$m features are given in Table~\ref{icetab}.
Meyer et al.~(\cite{meyer}) have proposed that the ice column density
correlates better with the ratio of mass-loss rate to luminosity ($\dot{M}/L$)
than with $\dot{M}$ alone. 
Adopting reasonable estimates (based on values in
the literature) for these parameters, our results support the relation between 
ice column density and $\log{\dot{M}/L}$ proposed by Meyer et al (see their
Fig.~3). However, given the uncertainties in both parameters, and that the
luminosity changes significantly with the variability phase, the
relationship should be treated with some caution.

OH32.8 shows the 3.1-$\mu$m ice band, and an 11-$\mu$m absorption  feature
in the
wing of the silicate absorption feature (Roche \& Aitken \cite{roche})  which
was attributed to the libration mode of water ice. 
Justtanont \& Tielens (\cite{justtiel}) were able to model ground-based
and
IRAS  observations of this source using silicate grains with water-ice mantles,
which give a much broader 10-$\mu$m absorption feature than do bare silicate
grains.

The broad 11-$\mu$m feature is clearly seen in the five sources with strong
10-$\mu$m absorption (Fig.~\ref{fig1}). 
It appears strongly in OH26.5, OH104.9, 
OH127.8 and OH32.8, and as an inflection near 11.5~$\mu$m in AFGL~5379.
The
contribution of this feature to the overall 10-$\mu$m absorption profile
therefore does not appear to
correlate fully with the presence of the other water ice bands:
AFGL~5379 shows strong far-IR ice emission and 3.1-$\mu$m absorption, but 
only weak 11-$\mu$m absorption,
while OH104.9 shows strong 11-$\mu$m absorption but has weak or absent far-IR
and 3.1-$\mu$m features.

Smith \& Herman (\cite{smith})  found an 11-$\mu$m absorption feature in
the spectrum of 
another OH/IR star,  OH138.0+7.3, which does not show any ice absorption at
3.1~$\mu$m. Since the 3.1-$\mu$m stretching mode is intrinsically stronger than
the libration mode, Smith \& Herman concluded that the 11-$\mu$m feature
observed towards OH138.0 is  not produced by water ice, and suggested that  it
is due to partially-crystalline silicates. Another possibility is that spectra
like that of OH138.0 are the absorption counterpart of the Little-Marenin \&
Little (\cite{lml}) `Sil++' or `Broad' emission features,  which show an
emission component at $\sim$11~$\mu$m on the wing of the silicate feature.
These features have been ascribed to crystalline silicates or amorphous alumina
grains (see e.g. Sloan \& Price \cite{sloan}). 
Clearly, full radiative-transfer modelling would be useful to determine 
whether 
ice mantles can indeed explain the range of 11-$\mu$m features seen, or whether
other grain components are necessary.

The presence of strong water ice features in our spectra indicates that a
substantial amount of the H$_2$O in the circumstellar envelopes may be
 depleted into
the solid phase. This would decrease the amount of gas-phase H$_2$O 
(and photodissociated OH) in the outer regions of the circumstellar
envelope which can be
detected by maser and thermal emission, 
Water maser lines are observed to be relatively weaker in OH/IR stars than
in objects
with lower mass-loss rates (e.g. Likkel \cite{likkel}). Collisional quenching
due to the high densities in the inner parts of the outflow is thought to
suppress the maser action; our results indicate that depletion into the
solid (ice) phase may also play a role.

CO ice shows features near 4.7~$\mu$m (e.g. Chiar et al.~\cite{chiar}): these
are not seen in our spectra, but a broad absorption band around  4.3~$\mu$m,
due to gas-phase CO is seen. This band is significantly broader than the
4.27-$\mu$m CO$_2$ ice absorption  feature seen in molecular clouds (e.g.
de~Graauw et al.~\cite{degraauwco2}). We see no evidence of CO$_2$ ice at
4.27~$\mu$m. CO$_2$ ice shows another strong feature at 15~$\mu$m; our spectra
show some structure near this wavelength (Fig.~\ref{ovcrystal}), 
but this may be an artefact of
the instrument or data-reduction process.

\subsection{Silicates}

For these objects, most of the energy radiated by 
the central star is absorbed by the circumstellar
dust shell and re-radiated as thermal emission, predominantly by amorphous
silicates. Amorphous silicates have strong features at 10 and 18 $\mu$m.
The current objects can be ordered by increasing optical depth of those 
features, which agrees with ordering by increasing wavelength of the
peak of the SED. When the 10-$\mu$m feature becomes optically thick,
the energy from the central star must be re-radiated at even longer
wavelengths, therefore the peak position shifts towards 30 $\mu$m 
(see Fig.~\ref{fig1}).
It is believed that the low mass loss rate Miras evolve into 
high mass loss rate OH/IR stars, while the IRAS colors indicate that the
peak of the dust emission shifts towards longer wavelengths  
(van der Veen \& Habing \cite{vanderveen}). 
With increasing mass loss rate 
the characteristic density in the wind will increase,
increasing the optical depth towards the central star. This evolution
can be found in the oxygen-rich AGB stars in our sample. 

\begin{figure*}
  \resizebox{\hsize}{!}{\includegraphics{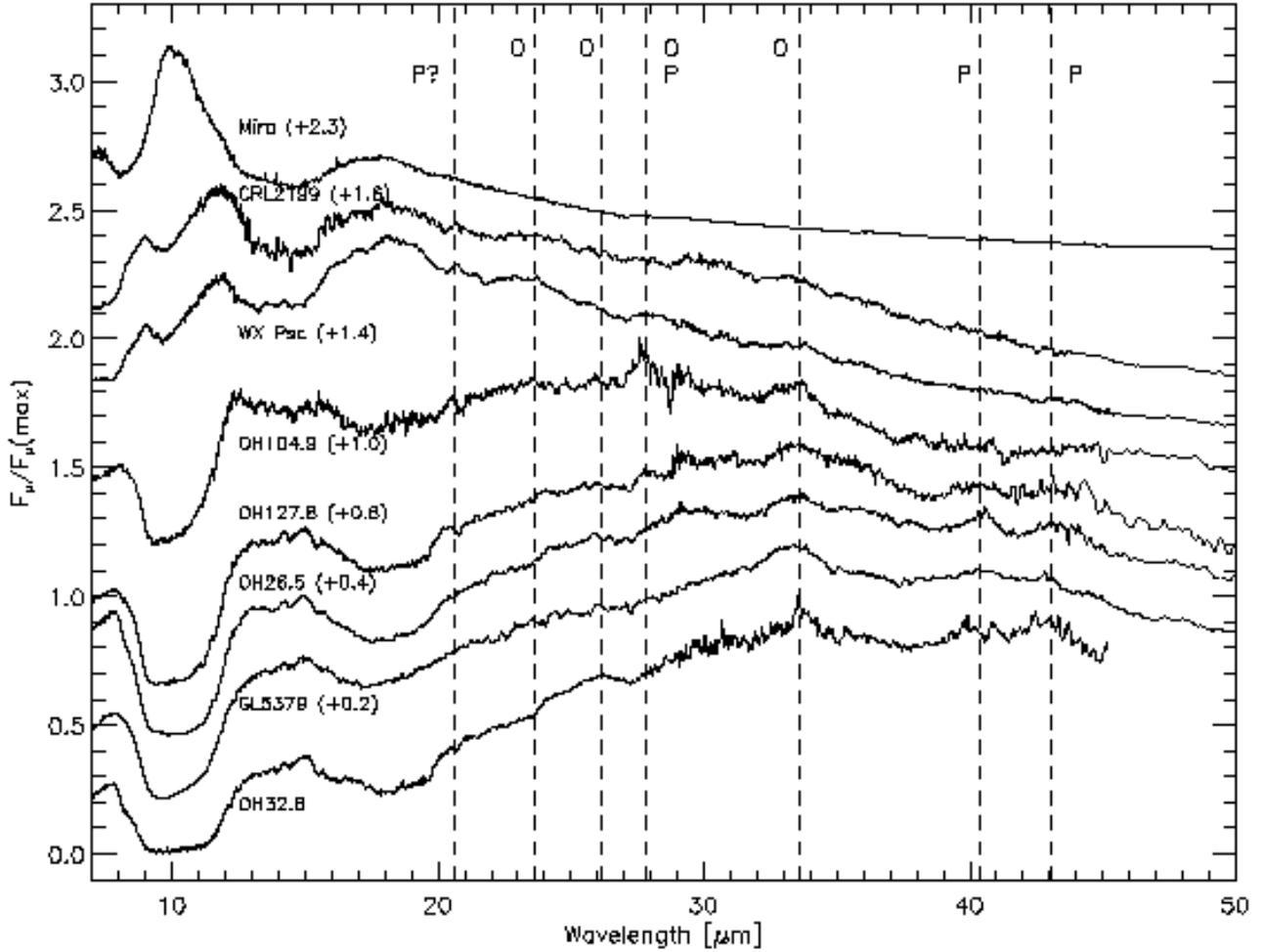}}
  \hfill
 \caption{Overview of the crystalline silicates. Intensities are plotted
in $F_\nu$ units, and are normalized by dividing by the maximum intensity.
The spectra are ordered by position of the peak in the determined continuum,
according to Fig.~\ref{fig1}, and shifted for clarity. The dotted lines 
indicate the wavelengths where crystalline silicate features are found in 
two or more
of the plotted spectra, see text. The crystalline species are indicated
by a P (pyroxene) or O (olivine); see Table~\ref{xtab} for details.}
    \label{ovcrystal}
\end{figure*}

In Fig.~\ref{ovcrystal} 
the objects are plotted in the same order as in Fig.~\ref{fig1}, 
however, the intensities are now in $F_\nu$ units, and normalized with 
respect to the measured maximum intensity. The 10-$\mu$m
feature of Mira is completely in emission; for WX~Psc and CRL~2199, the
10-$\mu$m feature is partially self-absorbed. For the OH/IR stars 
OH104.9, OH127.8, OH26.5, AFGL~5379 and OH32.8, the 10-$\mu$m
feature is completely in absorption, with optical depths 
ranging from 1.1 for OH104.9 to 3.6 for OH32.8. A detailed
analysis of the amorphous silicate features will be presented in a future
paper (Kemper et al, in preparation).
As the optical depth increases, some structure becomes  apparent 
in the 20--45 $\mu$m region.

\begin{table}
\caption{Crystalline silicate features in the spectra of our
programme stars. The features are detected by close examination of the
spectra.
Enstatite (pyroxene) is indicated by a P, forsterite (olivine) by an O.
If the feature is found in emission, it is indicated by a Y+, and by a
Y$-$ when the feature is found in absorption.}
\label{xtab}
\begin{tabular}{@{}l@{\hspace{1mm}}c@{\hspace{2mm}}c@{\hspace{2mm}}c@{\hspace{2mm}}c@{\hspace{2mm}}c@{\hspace{2mm}}c@{\hspace{2mm}}c@{\hspace{2mm}}c}
$\lambda$($\mu$m)&20.6 & 23.6 &26.1 &27.8 &33.0-33.6 &40.4 &43.1 &69\\
Species          & P?  & O    & O   & O,P & O,P      & P   & P &O\\
\hline
Mira             &N    &N     &N    &N    & N        & N   & N &N\\
CRL~2199         &Y+   &N     &N    &N    & Y+       & Y+  & N &N\\
WX~Psc           &Y+   &Y+    &N    &Y+   & Y+       & N   & N &N\\
OH104            &Y+   &Y+    &Y+   &Y+   & Y+       & N   & N &N\\
OH127            &Y$-$?&N     &Y+   &Y+   & Y+       & Y+  & Y+ & Y+\\
OH26             &Y$-$ &Y$-$  &Y+   &N    & Y+       & Y+  & Y+ & Y+\\
AFGL~5379        &N    &N     &Y+   &N    & Y+       & Y+  & Y+ & Y+\\
OH32             &Y$-$ &Y$-$  &Y+   &N    & Y+       & Y+  & Y+ & Y+\\
\hline
\end{tabular}
\end{table}

In the lower mass-loss rate
objects, strong emission features due to amorphous silicates
are present, but there are no obvious narrow features at $\lambda > 20$ $\mu$m.
For the redder objects, we find that the amorphous silicates features 
are \mbox{(self-)absorbed}, and that some structure is apparent at wavelengths
 $\lambda > 20$ $\mu$m.
These narrow features can be identified as crystalline silicates, both
olivines and pyroxenes (Waters et al.~\cite{waters96}). 
The identifications are based on laboratory spectra of
crystalline silicates (J\"ager et al.~\cite{jaeger}; 
Koike et al.~\cite{koike}, Koike \& Shibai \cite{koike98}) 
and similar bands seen in other objects on which
detailed studies have been performed, i.e.~AFGL~4106 
(Molster et al.~\cite{molster}) and to HD~45677 (Voors \cite{voors}).  
The dashed lines in Fig.~\ref{ovcrystal} represent the position
of some important crystalline silicate complexes, which are listed 
in Table~\ref{xtab}. These crystalline silicate features
are found in emission at the longest wavelengths but sometimes in absorption at
somewhat shorter wavelengths, for example the 23.6~$\mu$m olivine feature in
OH32.8 and in 
OH26.5 (see Fig.~\ref{ovcrystal}). The OH/IR stars presented here are the
only objects known to exhibit crystalline silicates in absorption outside
the 8-13 $\mu$m wavelength region. For OH32.8 this was already reported
by Waters \& Molster (\cite{wamo}) in comparison with AFGL~4106. 
The presence of the most important crystalline silicate features is indicated
in Table \ref{xtab} for all the sources in our sample. Note that those
features are detected by close examination of the spectrum; not all features
are visible in the overview figures presented in this paper.
Detailed modelling of the wealth of crystalline silicate
features shown by the individual objects is deferred to a future paper.

The crystalline silicate features tend to appear in those sources having 
greater 
optical depth at 10 $\mu$m. However, the sharpness of the crystalline
silicate features shows a large variation. 
The sharpness is expected to be determined by properties 
of the crystalline
silicates, such as the presence of impurities, the shape of the dust grains,
and irregularities in the lattice structure.
The spectrum of AFGL~5379 does not show the sharp crystalline peaks found 
in the spectra of the other OH/IR stars 
and in the spectra of AFGL~4106 and HD~45677, but the shapes of its
features do resemble the laboratory spectra of crystalline silicates
(J\"ager et al.~\cite{jaeger}). This suggests that the dust grains
around these objects exhibit differences
in the properties of the lattice structure, such as impurities, holes,
and edge effects due to grain size.
The appearance of the crystalline silicate emission features in the
redder sources in our sample confirms the relationship between dust
crystallinity and envelope colour temperature, and hence mass-loss rate,
identified by Waters et al.~(\cite{waters96}). Specifically, there seems
to be a threshold value for the mass loss rate above which the crystalline
silicate features appear in the spectrum. However, above this threshold
value, the strength and 
width of the features seem to be uncorrelated to the mass loss rate. 

\begin{figure}
  \resizebox{\hsize}{!}{\includegraphics{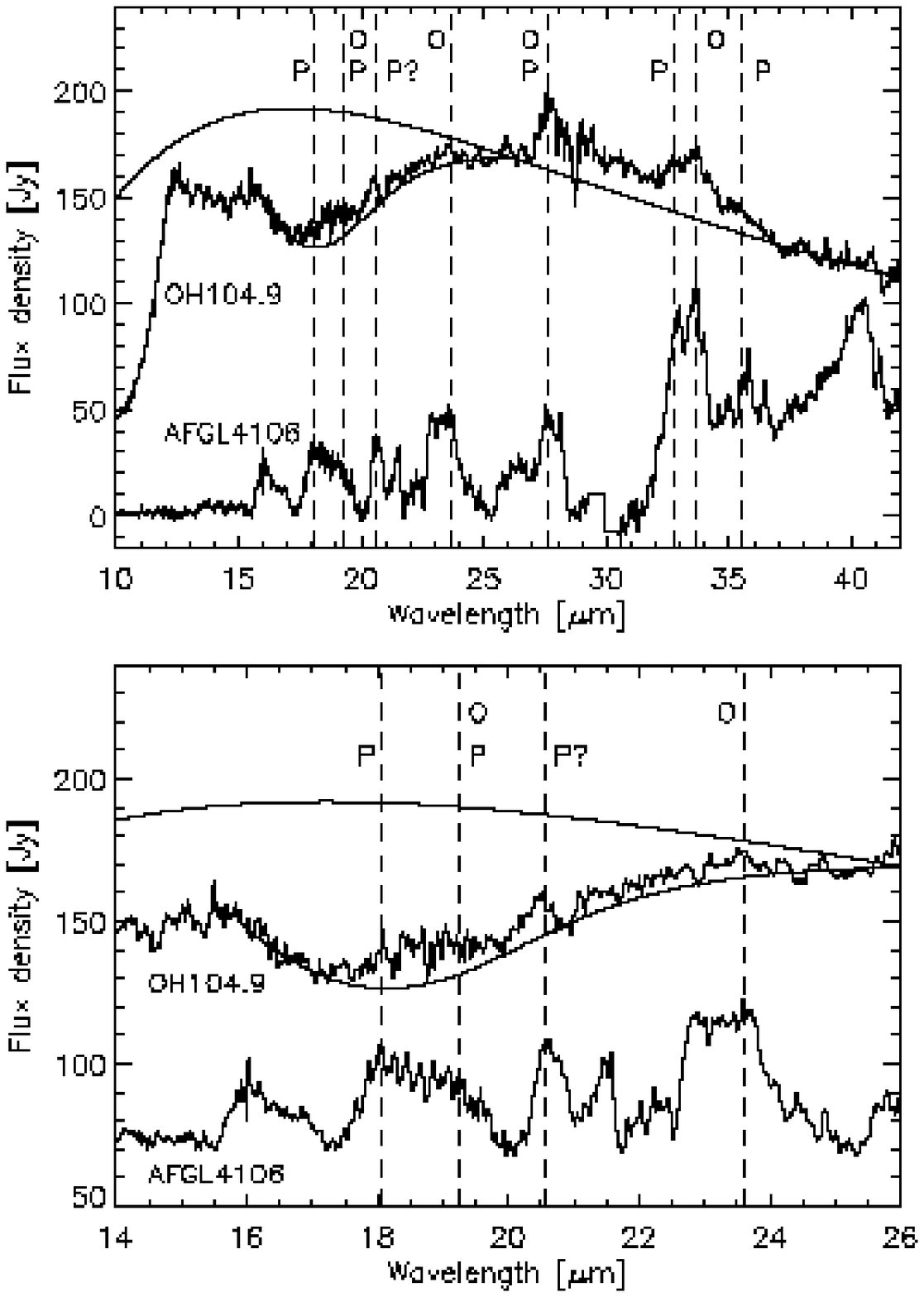}}
    \caption{The upper panel shows the spectrum of OH104.9 (upper 
spectrum), together with its continuum fit (solid line). A fit to the
18-$\mu$m absorption feature is also plotted. The lower spectrum in the upper
panel is the continuum subtracted spectrum of AFGL~4106 (Molster et 
al.~\cite{molster}). The dashed lines indicate a 
correlation between crystalline
silicate emission features in both spectra.
The lower panel is a close up of the upper panel, for the wavelength region
14 -- 26 $\mu$m; the spectrum of AFGL~4106 is shifted upward for clarity. The 
crystalline species is indicated on the plot by the
symbols P (pyroxene) and O (olivine).}
    \label{oh104}
\end{figure}

Fig.~\ref{oh104} presents a more detailed overview of the 
location of the crystalline silicate features.
In the upper panel, the upper spectrum is that
of OH104.9. The solid line represents the continuum fit obtained
using the method described above. 
At longer wavelengths the spectrum
shows emission features superposed on the  pseudo-continuum. 
The 18-$\mu$m \emph{absorption} feature
extends to  $\lambda \approx 26$ $\mu$m.
However, as Fig.~\ref{oh104} clearly shows for both
plots, there are crystalline \emph{emission} 
features present within this wavelength
region,
in particular, the 17.5--20 $\mu$m complex.
For reference, the continuum-subtracted spectrum of AFGL~4106 
(Molster et al.~\cite{molster}) 
is also plotted. 
The 20.6 $\mu$m emission feature is 
probably related to crystalline silicates, although it is not yet
identified (Molster et al.~\cite{molster}; Voors \cite{voors}).
These OH/IR star spectra are the first to show 
crystalline silicates in emission simultaneously with amorphous silicates in 
absorption in the same wavelength region.

The presence of crystalline emission features in the spectrum of OH104.9, 
at wavelengths where the amorphous dust
component is still in absorption, implies that the crystalline silicate dust 
must have a different spatial distribution than the amorphous silicate
dust.
We consider two possible geometries: spherical and axi-symmetrical.

For the case of a spherically symmetric
distribution, the crystalline dust can have a different radial
distribution and be
located further out in the envelope than the amorphous dust. 
The SWS and LWS beam sizes are much larger than the angular size 
of the dust shells of these OH/IR stars, so the amorphous
silicate absorption can originate from the
entire dust shell, while the crystalline silicate emission can arise from the
cool outer layers of the dust shell, where the material is optically thin. 
If the crystalline silicates are located further out, 
we can conclude that the crystalline
and amorphous dust has not formed at the same time, but that the
amorphous dust annealed as it moved away from the star.
This could
imply that lower mass loss rate Mira variables could in principle be able
to form crystalline material
as well, but that we cannot detect it because the column densities are not
high enough.
However, given that higher temperatures are required for annealing
amorphous silicates into crystalline silicates than are required for the
formation of amorphous silicates themselves, it does seem unlikely that
crystalline silicate grains could be formed further out in an outflow than
amorphous silicate grains. One possibility is that some small fraction of
the particles which formed in an outflow (perhaps the smallest particles)
immediately annealed into crystalline silicates in the inner, hottest
regions. The large total column density of amorphous silicate grains would
lead to net absorption in the 10- and 18-um bands and thus obscure these
hot crystalline silicates.
When these particles are cooled while flowing outwards, the
strong emission features of crystalline silicates at wavelengths longer
than $\sim$~15~$\mu$m can be seen  superposed on  the
amorphous dust features.

An alternative scenario for the observed behaviour is that OH/IR stars
possess a dust disk, in addition to a more spherically symmetric outflow.
Two alternatives suggest themselves:

\begin{itemize}
\item  Crystalline silicates that have formed and moved out from the inner
regions of the outflow produce emission bands that are seen superposed    
against the continuum emission and amorphous silicate features that arise
from the disk. This would require that those OH/IR stars that have the
largest 10-$\mu$m amorphous silicate optical depths are the ones whose
disks are seen most nearly edge-on.

\item The crystalline silicates are located in the disk, while the amorphous
silicates are located mainly in the outflow and thus have a more
spherically symmetric distribution, with the amorphous silicate absorption
features arising from optically thick lines of sight towards the central 
star. Depending on the inclination angle of the system, the crystalline  
silicate features can be seen in emission. When the disk is viewed
face-on, the crystalline silicate features would be optically thin.
Radiation from most of the disk surface reaches the observer via lines of
sight which pass through only the outer regions of the spherical dust  
shell, where the amorphous material is not optically thick.
\end{itemize}

Two mechanisms can be invoked to explain the high abundance of crystalline
silicates in a disk. Below the so-called glass temperature, only
amorphous silicates condense, while crystalline silicates can form at
temperatures greater than the glass temperature. At the high densities
occurring in a disk, condensation of silicates will be able to proceed
at higher temperatures than usual (Gail \& Sedlmayr \cite{gail})
and the temperature   
range in which it is possible to condense crystalline silicates is thus  
broadened. In our current sample there is empirical evidence for a
correlation between crystallinity and density. 
Second, if there is amorphous material present in the disk, this can be  
transformed into crystalline material by annealing. In order to allow the
annealing process of the silicates to proceed, the dust-forming region   
should not cool too rapidly. An orbiting (or slowly outflowing) disk
provides the required stability and keeps the amorphous silicates
relatively close to the central star for a sufficiently long time for the
annealing into crystalline material to occur. Both the above mechanisms   
provide circumstances that allow the formation of crystalline silicates,
which would not be the case if the stellar wind removes the newly-formed 
silicates at the outflow velocity, such that they rapidly cool.

In order to study the spatial distribution of both dust components in more
detail, spectral mapping of the OH/IR stars is necessary. Then we may be
able to put constraints on the spatial distribution, and determine the
annealing time, of the crystalline dust, using travel time and stability
arguments. Using laboratory data on annealing time scales, we may be able
to determine physical parameters such as temperature and density in the
circumstellar dust shell, thus helping to clarify the AGB mass loss phase
of stellar evolution. Recent high-resolution imaging of the dust around
evolved stars, such as Mira (Lopez et al.~\cite{lopez}) and VY~CMa (Monnier
et al.~\cite{monnier99}), has shown that substantial 
inhomogeneities exist in the
dusty outflows; our spectra suggest that the dust around OH/IR stars may
also show complex morphologies. 

For any geometry of the circumstellar shell, the amorphous--crystalline
volume ratio could also be affected by grain-grain collisions.  Such
collisions have long been recognised as important processes for the
evolution of grains in circumstellar envelopes (see e.g.~Biermann \&
Harwit \cite{bier}) and in the interstellar medium (Jones et 
al.~\cite{jones}). 
The shock wave driven into the two grains by the collision can lead to
vaporization, the formation of high pressure phases, melting, annealing,
and shattering depending on the pressures involved (cf., Tielens et 
al.~\cite{tmsh}).  Experiments show that at a relative collision velocity of
\mbox{$\sim$1~km s$^{-1}$}, mechanical 
effects (shattering, crater formation) become
important.  Thermal effects such as crystallization, involving the
intergranular nucleation of new, strain--free grains in a previously
highly deformed matrix start at somewhat higher velocities (\mbox{$\sim$ 
5 km s$^{-1}$}; 700 kbar) and are 
never very pervasive.  Above \mbox{$\sim$ 7 km s$^{-1}$} (1 Mbar),
melting followed by rapid quenching leads to glass formation (Bauer
\cite{bauer}; Schaal et al.~\cite{schaal}).  Thus, these experiments imply that
crystallization only occurs over a very narrow collision velocity range
and is not very efficient.  Of course, if the projectile/target size
ratio is large, even high velocity impacts will lead to a small volume
fraction of the target grain passing through the regime where
recrystallization can occur when the shock wave expands. 

Given the grain velocity profile in AGB outflows (Habing et 
al.~\cite{habingea}),
potentially crystallizing grain-grain collisions will be largely confined
to the acceleration zone at the base of the flow.  An important
objection against annealing through grain-grain collisions is that the
cooling time scales of dust grains are very short compared to the
annealing time scales.  Therefore the silicate dust grains solidify in
the amorphous state (Molster et al.~\cite{molnat}).  The annealing
process can be described as a three dimensional random walk diffusion
process on a cubic lattice (Gail \cite{hpgail}), which provides an
estimate of the annealing time scale as a function of dust temperature. 
At a \mbox{$T_{\rm d}$ = 2000~K} 
the annealing time scale is \mbox{$\sim$ 1~s} and
strongly increases for lower temperatures.  The cooling time scale can
be derived under the assumption that the power emitted by the dust grain 
is given by the Planck function, 
modified by the Planck mean of the absorption efficiency
$Q_{\rm{abs}}$. By
comparing the emitted power to the internal heat of the grain one finds for the
cooling time scales \mbox{$\tau_{\rm cool}$ $\sim$ 10$^{-3}$~s} 
for \mbox{$T_{\rm d}$ = 2000~K}
and \mbox{$\tau_{\rm{cool}}$ $\sim$ 0.02~s} for \mbox{$T_{\rm d}$ = 800~K}.  
In agreement
with the experiments, only a small fraction of crystallized material is
expected due to the grain-grain collisional shock loading.  Final
assessment of the viability of this mechanism for the formation of
crystalline silicates in AGB outflows has to await detailed modeling of
such grain-grain collisions in circumstellar outflows (Kemper et al.~in 
preparation).  Finally, we note that the spectrum of \mbox{$\beta$ Pic}, 
which is due to a dust size distribution that is 
surely collisionally dominated, shows little evidence for
crystalline silicates. Although the 10~$\mu$m spectrum of \mbox{$\beta$ Pic} 
(Knacke et al.~\cite{knacke}, Aitken et al.~\cite{aitken}) may
show some spectral structure resembling that of solar system comets --
characteristic for silicate minerals -- the longer wavelength bands so
prominent in cometary spectra are completely absent in this source
(Pantin et al.~\cite{malfbpic}). 
While this may at first sight argue against this
mechanism, the collisional velocities in this system may be much lower
and predominantly lead to shattering. 

\section{Conclusions}
We have presented the complete SWS/LWS spectra for seven oxygen-rich evolved
stars, together with the SWS spectrum of an eighth source. 
For the OH/IR stars, which have optically thick dust shells, essentially
all of the stellar luminosity is radiated in the wavelength region covered by
ISO. Emission features of crystalline silicates are seen longwards of
15~$\mu$m in the dustier objects. Some of these emission features lie within
the 20-$\mu$m absorption feature of amorphous silicate, suggesting that the
crystalline and amorphous components have different spatial distributions.
The dust shells of these sources are sufficiently cool for abundant water ice
to form, as indicated by the near-IR absorption and far-IR emission features of
crystalline ice.

In a future paper (Kemper et al., in preparation) 
we will present some of the further
analysis and modelling required to determine the physical conditions
and processes which give rise to these very rich spectra.

\begin{acknowledgements}
We thank Tanya Lim for making available the spectral synthesis routine,
and the referee, H.~Habing, for constructive comments. 
FK and LBFMW acknowledge financial support from NWO Pionier grant number
616-78-333, and from an NWO Spinoza grant number 08-0 to E.P.J.~van den
Heuvel.
\end{acknowledgements}

\end{document}